\documentclass[conference]{IEEEtran}
\usepackage{cite}
\usepackage{tabularx}
\usepackage{tabulary}
\usepackage{array}
\usepackage{booktabs}
\usepackage{verbatim}
\usepackage{amsmath}
\usepackage{listings}
\usepackage{xcolor} 
\usepackage{makecell} 						
\usepackage{csquotes}
\usepackage{multirow}
\usepackage{graphicx}
\usepackage{stix}
\usepackage[bookmarks=false]{hyperref}

\definecolor{light-gray}{gray}{0.95}

\DeclareRobustCommand{\aasIndent}{\text{\reflectbox{$\carriagereturn$}}}

\begin{document}

\bstctlcite{IEEEexample:BSTcontrol} 

\title{Toward a Mapping of Capability and Skill Models using Asset Administration Shells and Ontologies}

\author{
    \IEEEauthorblockN{
        Luis Miguel Vieira da Silva\IEEEauthorrefmark{1},
        Aljosha Köcher\IEEEauthorrefmark{1},
        Milapji Singh Gill\IEEEauthorrefmark{1},
        Marco Weiss\IEEEauthorrefmark{2}
        and~Alexander Fay\IEEEauthorrefmark{1}%
    }
    
    \IEEEauthorblockA{
        \IEEEauthorrefmark{1}Institute of Automation Technology\\
        Helmut Schmidt University, Hamburg, Germany\\
        {\tt\small \{miguel.vieira, aljosha.koecher, milapji.gill, alexander.fay\}@hsu-hh.de}\\
        \IEEEauthorrefmark{2}Institute of Maintenance, Repair and Overhaul\\
        German Aerospace Center (DLR), Hamburg, Germany\\
    }
}

\maketitle

\begin{abstract}

In order to react efficiently to changes in production, resources and their functions must be integrated into plants in accordance with the plug and produce principle. 
In this context, research on so-called capabilities and skills has shown promise. 
However, there are currently two incompatible approaches to modeling capabilities and skills.
On the one hand, formal descriptions using ontologies have been developed. 
On the other hand, there are efforts to standardize submodels of the Asset Administration Shell (AAS) for this purpose. 
In this paper, we present ongoing research to connect these two incompatible modeling approaches. 
Both models are analyzed to identify comparable as well as dissimilar model elements. 
Subsequently, we present a concept for a bidirectional mapping between AAS submodels and a capability and skill ontology. For this purpose, two unidirectional, declarative mappings are applied that implement transformations from one modeling approach to the other - and vice versa.
\end{abstract}

\begin{IEEEkeywords}
Capabilities, Skills, Ontologies, AAS, Mapping, RML, RDFex, Transformation, Semantic Web, OWL, RDF
\end{IEEEkeywords}

\section{Introduction} \label{introduction} 

Paradigms such as plug and produce are a promising way to achieve customer-specific production with a multitude of product variants. 
A central aspect of this approach is to reduce the effort required to integrate new resources and their functions. 
For plug and produce, a machine-interpretable information model of resources including individual mechatronic modules and their functions is indispensable \cite{CFL_Dynamicskillallocationmethodology_2013}. 

Current research in this area aims to create models for so-called \emph{capabilities} and \emph{skills}. In this context, a capability is defined as an \enquote{implementation-independent specification of a function}, whereas a skill is defined to be the executable implementation of said function \cite{KBH+_AReferenceModelfor_15.09.2022b}.

To this date, different approaches to modeling capabilities and skills have been developed, of which the two most recent are discussed in this paper.
On the one hand, there are formal models using ontologies \cite{ JSL_FormalResourceandCapability_2016, KHV+_AFormalCapabilityand_9820209112020}. 
On the other hand, there are approaches using the Asset Administration Shell (AAS) \cite{Sidorenko.2021, HDM_AASCapabilityBasedOperationand_9720219102021}. 
For the latter, the \emph{Industrial Digital Twin Association (IDTA)} coordinates efforts to standardize AAS submodels. 

Both of these modeling approaches offer certain advantages and disadvantages, depending on the perspective and requirements derived from a given use case.
We aim to develop an automated and bidirectional transformation between capability and skill models implemented using AASs and ontologies. 
This transformation will allow equipment described with different model types to be used in an interoperable way. Furthermore, it supports switching between model types in order to make use of the advantages and tools developed for each model type.

The contribution of this paper is twofold: First, a comparison of capability and skill models with AASs as well as ontologies is carried out in order to identify similarities and differences. 
Based on this analysis, an approach for a bidirectional mapping between these capability and skill modeling approaches is introduced. 
Accordingly, the structure of the paper is as follows: In Section~\ref{relatedWork}, related work is presented followed by the aforementioned contribution in Section~\ref{contribution}. 
A summary and outlook in Section~\ref{summaryOutlook} concludes this paper.

\section{Related Work} \label{relatedWork}

A working group of \emph{Plattform Industrie 4.0} develops shared definitions and an information model of capabilities and skills. As an output of this working group, an abstract reference model is presented in \cite{KBH+_AReferenceModelfor_15.09.2022b}.
This model is currently implemented with two different technologies. 

On the one hand, there are modeling approaches using ontologies \cite{JSL_FormalResourceandCapability_2016, KHV+_AFormalCapabilityand_9820209112020, DKF_Acapabilityandskill_2023}. The capability and skill ontology\footnotemark{} presented in \cite{KHV+_AFormalCapabilityand_9820209112020} has been continuously improved and is built on a three-layer ontology architecture. It extends the reference model of \cite{KBH+_AReferenceModelfor_15.09.2022b} with manufacturing-specific details through reusable ontology design patterns, which are based on standards. There is also an extension of the model presented in \cite{KBH+_AReferenceModelfor_15.09.2022b} for capabilities and skills of autonomous robots \cite{DKF_Acapabilityandskill_2023}.

\footnotetext{Ontology available at \url{https://github.com/aljoshakoecher/caskman}}

On the other hand, there are modeling approaches for capabilities and skills using the AAS. In \cite{Sidorenko.2021}, Sidorenko et al. present an OPC UA skill model that is intended to be used by proactive AAS in a semantic interaction protocol. Hereby, decentralized approaches to production control should be enabled by allowing Industry 4.0 components to communicate directly with each other as well as to mutually use skills. 
In \cite{HDM_AASCapabilityBasedOperationand_9720219102021}, a model-driven engineering approach is presented that uses an AAS model of capabilities in order to achieve interoperable components and a flexible production line.

To standardize developments of the AAS in this context, working groups of the IDTA create submodel templates and specifications for capabilities \cite{CapabilitySubmodel} and skills \cite{ControlComponentType, ControlComponentInstance}.

Beden et al. \cite{BCB_SemanticAssetAdministrationShells_2021} conducted a literature survey of different research approaches trying to add explicit semantics to multiple types of data models used in the context of Industrie 4.0.
They argue that the AAS lacks a formal specification and that many papers propose fragmented models or focus on specific use cases without reusability in mind. 
Furthermore Beden et al. state that research approaches trying to add formal semantics (e.g., through mappings) are still in early stages but consider such approaches to be an important challenge. \cite{BCB_SemanticAssetAdministrationShells_2021} 

In \cite{BaMa_TheSemanticAssetAdministration_2019}, Bader and Maleshkova present an RDF model for semantic AASs and a unidirectional mapping to convert an AAS into RDF representation. 
In order to infer implicit information and validate the automatically generated RDF models, reasoning axioms and constraints are presented, too.
Similar to \cite{BaMa_TheSemanticAssetAdministration_2019}, our approach also makes use of mapping rules to convert the JSON representation of an AAS into an ontology. 
However, while Bader and Maleshkova generate an ontology strictly following the AAS specifications on meta-model level, we transform AAS information into the existing capability and skill ontology of \cite{KHV+_AFormalCapabilityand_9820209112020}. This ontology adheres to the reference model of \cite{KBH+_AReferenceModelfor_15.09.2022b}, but additionally benefits from a tool ecosystem with tools for model generation and production execution.


As a summary of related publications, it can be said that there are currently two incompatible modeling approaches for capabilities and skills. While there is a unidirectional transformation from AAS to RDF in general, there is no detailed transformation for capabilities and skills until now. Furthermore, a bidirectional transformation does not yet exist.

\section{Contribution} \label{contribution} 

This section contains a comparison of the AAS and ontology modeling approaches of capabilities and skills. The comparison is based on the reference model published in \cite{KBH+_AReferenceModelfor_15.09.2022b}. We contrast the AAS submodels \emph{Capability} \cite{CapabilitySubmodel} and \emph{Control Component} (type and instance) \cite{ControlComponentType, ControlComponentInstance} with the \emph{CaSkMan} ontology initially presented in \cite{KHV+_AFormalCapabilityand_9820209112020}.
Based on this comparison, a bidirectional mapping approach is presented.

\subsection{Comparison and Analysis}\label{Comparison}


\begin{table*}[h!]
    \centering
    \caption{Comparison of the most important capability and skill model elements of the examined ontology and the AAS submodels.\\ Ontology elements are represented in Description Logic syntax, while the hierarchical nesting is shown for AAS elements.\\ \normalfont{
    (SM: Submodel, \quad SMC: SubmodelCollection, \quad Cap: Capability, \quad Rel: Relation, \quad Prop: Property, \quad Ref: Reference)}}
    \label{tab:comparison}
    \renewcommand*{\arraystretch}{2}
    \begin{tabularx}{\linewidth}{l c c}
    \toprule
        \thead[l]{\textbf{Model Element}} &
        \thead[l]{\textbf{Ontology} \\ \emph{CaSkMan}} & 
        \thead[l]{\textbf{AAS} \\ SM \emph{Capability} \& SM \emph{Control Component}} \\
        \midrule
        \textbf{Capability} & \makecell[l]{$\textrm{Capability}$ \\ $\textrm{ProvidedCapability} \equiv \textrm{Capability} \sqcap \exists \textrm{provides}^{-1}.\textrm{Resource}$} & \makecell[l]{SMC CapabilitySet \\ 
        \aasIndent{} SMC CapabilityContainer $\rightarrow$ Cap Capability} \\
        \quad Property & \makecell[l]{$\textrm{Property} \sqcap \exists \textrm{isSpecifiedBy}^{-1}.\textrm{Capability}$} & \makecell[l]{SMC PropertySet \\ \aasIndent{} SMC PropertyContainer $\rightarrow$ Prop Property} \\
        \quad Capability constraint & \makecell[l]{$\textrm{CapabilityConstraint} \sqcap \exists \textrm{isRestrictedBy}^{-1}.\textrm{Capability}$ \\ \qquad $\sqcap \; \exists \textrm{references}^{-1}.\textrm{Property}$} & \makecell[l]{SMC CapabilityRelationships \\ \aasIndent{} SMC ConditionContainer $\rightarrow$ Rel CapabilityCondition} \\ 
        \quad Process type & \makecell[l]{$\textrm{Process} \sqcap \exists \textrm{requires.Capability}$ \\ $\textrm{(ManufacturingProcess} \sqcup \textrm{HandlingProcess)} \sqsubseteq \textrm{Process}$} & \makecell[l]{-} \\
        \textbf{Skill} & \makecell[l]{$\textrm{Skill} \sqcap \exists \textrm{isRealizedBySkill}^{-1}.\textrm{Capability}$} & \makecell[l]{SMC Skills \quad\quad\quad \& \quad SMC CapabilityRelationships \\ \aasIndent{} SMC Skill \qquad\qquad \, \aasIndent{} Rel realizedBy} \\
        \quad Skill parameter & \makecell[l]{$\textrm{SkillVariable} \sqcap \exists \textrm{hasSkillVariable}^{-1}.\textrm{Skill}$ \\ \qquad $\sqcap \; \exists \textrm{isRealizedBySkillParameter}^{-1}.\textrm{Property}$} & \makecell[l]{SMC Parameters \quad \& \quad SMC PropertyRelationships \\ \aasIndent{} SMC Parameter \qquad \, \aasIndent{} Rel realizedBy} \\
        \quad State machine & \makecell[l]{$\textrm{StateMachine}  \sqcap \exists \textrm{behaviorConformsTo}^{-1}.\textrm{Skill}$} & \makecell[l]{SMC Modes \quad\quad \; \& \quad Prop Disabled \\ \aasIndent{} Prop Mode} \\
        \quad Skill interface & \makecell[l]{$\textrm{SkillInterface} \sqcap \exists \textrm{accessibleThrough}^{-1}.\textrm{Skill}$ \\ \qquad  $\sqcap \; \exists \textrm{exposesStateMachine.StateMachine}$} & \makecell[l]{SMC:Interfaces \quad \, \& \quad SMC Endpoints\\ \aasIndent{} Ref Interface \qquad\quad\; \aasIndent{} Ref Endpoint} \\         
        \bottomrule
    \end{tabularx}
\end{table*}

The model element considered first is \emph{capability}. In the ontology, capabilities are represented as individuals of the class \emph{Capability}, which has subclasses for required and provided capabilities. Provided capabilities are related to the providing resource via an object property \emph{provides}. 

To an AAS of a resource, offered capabilities can be assigned with a specific submodel of type \emph{Capability}. 
Individual capabilities can be added via a SubmodelElement (SME) \emph{Cap}. 
Each \emph{Cap} instance must reside in a SubmodelCollection (SMC) of type \emph{CapabilityContainer} that contains other capability-related information, such as properties or relationships to other elements. 
All capability containers of an asset are stored inside an SMC \emph{CapabilitySet}. The capability submodel does not explicitly distinguish between required and provided capabilities.   

In the ontology, properties of a capability are modeled  as individuals of the class \emph{Property} and linked to a capability with the object property \emph{isSpecifiedBy}. To formally express statements about properties (e.g., requirements), a data element according to IEC 61360 is used.
In an AAS, a SMC \emph{PropertySet} is located inside a capability container. Similar to the way capabilities are modeled, there is an SMC \emph{PropertyContainer} with a nested SME \emph{Prop} for each property. 

A capability can be restricted by constraints.
In the ontology, a constraint is defined as an individual of the class \emph{CapabilityConstraint} and additional mathematical expressions according to OpenMath. Every constraint is linked to its capability and the properties to which it refers via \emph{isRestrictedBy} and \emph{references}, respectively.
In the AAS submodel \emph{Capability}, there is an SMC \emph{ConditionContainer}, in which capability constraints can be specified via relations.
However, it is yet to be defined how constraints are formulated in detail. 

According to \cite{KBH+_AReferenceModelfor_15.09.2022b}, capabilities are linked to processes. To model the process linked to a capability in detail, the ontology contains process types according to DIN 8580 and VDI 2860. Furthermore, inputs and outputs of a process can be modeled according to VDI 3682. 
The AAS submodel \emph{Capability} does currently not allow to model the process linked to a capability in more detail. However, process types can be represented with a semanticId of a capability using a respective ECLASS classification. 
Nevertheless, it is possible to compose or decompose capabilities in both approaches. 

In order to model skills, the ontology defines a class \emph{Skill}. Individuals of this class are linked to a capability via \emph{isRealizedBySkill}. 
In the AAS, the submodels \emph{Control Component Type} and \emph{Control Component Instance} are used to specify skills. 
Skills are collected in the SMC \emph{Skills} and a single skill is modeled via the SMC \emph{Skill}. 

In the ontology, parameters of a skill are modeled as individuals of the class \emph{SkillParameter} and assigned to a skill via \emph{hasSkillVariable}.
In a \emph{Control Component}, parameters are represented via an SMC \emph{Parameter}, which resides in an SMC \emph{Parameters}, which in turn resides in an SMC \emph{Skill}. 
Both approaches allow type and values of parameters to be modeled. 
A skill parameter may realize a capability property. 
Therefore, the corresponding elements are linked using \emph{isRealizedBySkillParameter} in the ontology. 
In the AAS submodel \emph{Capability}, a property is linked to a related skill parameter via a Relationship element \emph{realizedBy} that is located in the SMC \emph{PropertyRelationShips} of the SMC \emph{PropertySet}. 

A skill is implemented with a state machine that controls the interaction with a skill (e.g., start or stop). 
The ontology models a state machine as an individual of the class \emph{StateMachine} and an ontology according to ISA 88, which is used to explicitly depict all states and transitions. 
A skill is then linked to its state machine via \emph{behaviorConformsTo}. 
In the AAS, a state machine is not explicitly modeled. Instead, different execution modes of a skill are modeled with an SMC \emph{Modes}. In addition, there is a Property element \emph{Disabled}, which indicates whether a skill is available. 

Each skill must have a skill interface that allows to interact with the skill by triggering transitions of the state machine or setting parameters \cite{KBH+_AReferenceModelfor_15.09.2022b}. 
In the ontology, an interface is represented as an individual of the class \emph{SkillInterface}, which is attached to a skill with the object property \emph{accessibleThrough}. 
Specific subclasses for different types of interfaces exist. 
A skill's interface exposes its state machine, methods to trigger transitions and parameters. 
When using an AAS, possible interfaces are specified in the SMC \emph{Interfaces} by individual ReferenceElements \emph{Interface} of the \emph{Control Component Type} submodel.
In the submodel \emph{Control Component Instance}, individual endpoints are defined in the SMC \emph{Endpoints} with a ReferenceElement \emph{Endpoint}.
The SMC \emph{Interfaces} in its current form lacks features that are needed to model skill interfaces with technologies such as OPC UA or HTTP. 
For example, it is unclear how endpoints are related to the individual skills and how operations and parameters are modeled.

In conclusion, both model approaches enable modeling of capabilities and skills, albeit on different levels of detail.
The ontological approach is currently more expressive and models all elements of the CSS model on a detail that allows production planning and execution. On the other hand, the submodels of the AAS are not yet specified on this level of detail. 
Table~\ref{tab:comparison} provides an overview of the comparison of the two model approaches.

\subsection{Mapping Concept}
\label{Mapping}

After this comparison, we now describe a bidirectional mapping approach that allows to transform between a model represented with the aforementioned AAS submodels $M_{AAS}$ and a model represented with the \emph{CaSkMan} ontology $M_{Onto}$.
For this purpose, the transformation from one AAS model element to the capability and skill ontology is first defined as

\begin{equation*}
  f_{A->O}: m_{AAS} \mapsto m_{Onto} \quad \forall m_{AAS} \in M_{AAS} \,.
\end{equation*}

This transformation is implemented in a declarative way using the RDF Mapping Language (RML). RML, first introduced in \cite{DVC+_RML:AGenericLanguage_2014b}, is a generic mapping language that can be utilized to express user-defined rules to transform information from heterogeneous data sources, e.g., XML or JSON into RDF.

\begin{lstlisting}[
caption={RML mapping rule to transform a capability from an AAS submodel to a capability individual in the ontology},
label={lst:rml-mapping}]
<#Capability-Mapping> a rr:TriplesMap;
 rml:logicalSource [
  rml:source "<Capability Submodel>.json";
  rml:referenceFormulation ql:JSONPath;
  rml:iterator "<path to all capabilities>" 
];

 rr:subjectMap [
  rr:template "http://..#{@idShort}";
  rr:class CSS:Capability
].
	
\end{lstlisting}

Listing~\ref{lst:rml-mapping} contains an example for an individual mapping rule that is used to map a capability modeled in an AAS into the ontology.
With \verb|rml:logicalSource|, the source file is specified and the elements to iterate over are expressed using JSONPath syntax. The example of Listing~\ref{lst:rml-mapping} iterates over all capability submodel elements in the given submodel. Please note that the actual iterator expression is too long to fit into this example, but all mappings are available on GitHub\footnote{https://github.com/hsu-aut/CSS-AAS-OWL}.
The \verb|rr:subjectMap| is applied for every iterator element, effectively creating an individual of the class \verb|CSS:Capability| with an IRI that makes use of the current capability's \emph{idShort}.
Besides one \verb|rr:subjectMap|, every mapping rule may have multiple \verb|rr:predicateObjectMap| elements to create relations of an individual. For the example of Listing~\ref{lst:rml-mapping}, a \verb|rr:predicateObjectMap| could be used to transform a capability's AAS comment into an \verb|rdfs:description| of the corresponding individual in the ontology.

\medskip

The opposite transformation, i.e., from an ontological capability and skill representation into an AAS is defined as

\begin{equation*}
  f_{O->A} = f_{A->O}^{-1} : m_{Onto} \mapsto m_{AAS} \quad \forall m_{Onto} \in M_{Onto} \,.
\end{equation*}

Since RML is a unidirectional mapping language, it cannot be used to implement $f_{O->A}$. Instead, we make use of RDFex, which was created as a counterpart to RML \cite{KMF_TowardaGenericMapping_2022}. 
RDFex has a syntax inspired by RML but allows writing mapping rules that target elements of an ontology and transform them into JSON or XML according to a user-specified schema.

\begin{lstlisting}[
caption={RDFex mapping rule to transform a capability individual from an ontology to a capability element in an AAS},
label={lst:rdfex-mapping}]
<#CapabilityMapping> a rdfex:DataMap;
 rdfex:ontologicalSource [
  rdfex:source "<Capability Ontology>.ttl";
  rdfex:sourceType rdfex:File;
  rdfex:queryLanguage rdfex:Sparql;
  rdfex:query '''
  PREFIX CSS: <http://www.w3id.org/hsu-aut/css#>
  SELECT ?cap (STRAFTER(STR(?cap), '#') AS ?capName)
  WHERE { ?cap a CSS:Capability }'''
 ];
 
 rdfex:targetFormat rdfex:JSON;
 rdfex:container "<path to CapabilitySet>";
 rdfex:snippet '<capability element with idShort defined by ?capName'.
\end{lstlisting}

The example of Listing~\ref{lst:rdfex-mapping} is the inverse transformation of Listing~\ref{lst:rml-mapping}. In this example, an ontology contained in a file represents the mapping's source.
A SPARQL query is used to select elements in the source ontology that are transformed with the current mapping rule. 
In this example, all capability individuals are selected and their local names are stored in the variable \verb|?capName|.
As our implementation of $f_{O->A}$ makes use of an AAS's JSON serialization, the \verb|targetFormat| is set to JSON.
In RDFex, a \verb|container| defines the parent element into which a mapping's output will be inserted. For this capability mapping, a search path in JSONPath syntax is defined that looks for the \emph{CapabilitySet} element in a given AAS.
And finally, the \verb|snippet| defines the shape of the actual output, taking into account variables of the SPARQL query. In this example, a complete SMC and capability SME is defined with the variable \verb|?capName| as the \verb|idShort| of the capability SME. A snippet is inserted into the container for each variable binding used.

\section{Summary And Outlook} \label{summaryOutlook} 
In this paper, a concept for a bidirectional mapping between capability and skill models in AASs and an ontology has been outlined. 
For this purpose, an ontology modeling approach was compared to AAS submodels before rule-based mappings were introduced using RML and RDFex.

With respect to future work, the following points are essential: As the analysis has shown, a complete mapping is not yet possible due to missing model elements in the existing AAS submodels. 
Therefore, the existing submodels need to be supplemented with more detailed concepts and relations in order to be fully usable for the capability and skill application.
 
We plan on extending our mapping rules and integrating the two types of rules into one application. This application will be embedded into our skill-based control platform so that both AAS and ontological capability and skill models can be used in manufacturing processes.
Additionally, the presented approach will also be evaluated in the experimental maintenance, repair and overhaul setting for aircrafts presented in \cite{WWW_MaSiMOAHybridExperimental_2021}.

\section*{Acknowledgment}
This research in the projects ProMoDi and RIVA is funded by dtec.bw – Digitalization and Technology Research Center of the Bundeswehr. dtec.bw is funded by the European Union – NextGenerationEU
 
\bibliographystyle{./bibliography/IEEEtran}
\bibliography{./bibliography/references} 

\end{document}